\documentclass[showpacs,showkeys,superscriptaddress,amsmath,amssymb,twocolumn,reprint,pra,10pt]{scrartcl}
\usepackage{amsmath,amssymb,graphicx,xcolor,cuted}
\usepackage{bm,hyperref}
\usepackage{epsfig,epstopdf}
\usepackage{mathrsfs,subfigure,dcolumn,enumitem}
\usepackage[left=0.5in,right=0.5in,top=0.5in,bottom=0.75in]{geometry}

\newcommand{\ket}[1]{|#1\rangle}
\newcommand{\be}{\begin{eqnarray}}
\newcommand{\ee}{\end{eqnarray}}

\begin{document}

\title{Exact $\mathcal{N}$-point function mapping between pairs of experiments with Markovian open quantum systems}

\author{Etienne Wamba$^{1,2,3,4}$, and Axel Pelster$^{4}$
	\\
	${}^{1}${\footnotesize \textit{Faculty of Engineering and Technology, University of Buea, P.O. Box 63 Buea, Cameroon,}}\\
	${}^{2}${\footnotesize \textit{African Institute for Mathematical Sciences, P.O. Box 608 Limbe, Cameroon, }}\\
	${}^{3}${\footnotesize \textit{International Center for Theoretical Physics, 34151 Trieste, Italy,}}\\
	${}^{4}${\footnotesize \textit{Fachbereich Physik and State Research Center OPTIMAS,}}\\{\footnotesize \textit{ Technische Universit\"at Kaiserslautern, 67663 Kaiserslautern, Germany}}\\
	{\footnotesize \textit{\ \textbf{E-mail}: wamba.etienne@gmail.com,~axel.pelster@physik.uni-kl.de }}}

\date{\small \today \vspace*{-2cm}}

\maketitle

\begin{strip}
	\section*{Abstract}
%
We formulate an exact spacetime mapping between the $\mathcal{N}$-point correlation functions of two different experiments with open quantum gases. Our formalism extends a quantum-field mapping result for closed systems [Phys. Rev. A \textbf{94}, 043628 (2016)] to the general case of open quantum systems with Markovian property. For this, we consider an open many-body system consisting of a $D$-dimensional quantum gas of bosons or fermions that interacts with a bath under Born-Markov approximation and evolves according to a Lindblad master equation in a regime of loss or gain. Invoking the independence of expectation values on pictures of quantum mechanics and using the quantum fields that describe the gas dynamics, we derive the Heisenberg evolution of any arbitrary  $\mathcal{N}$-point function of the system in the regime when the Lindblad generators feature a loss or a gain. Our quantum field mapping for closed quantum systems is rewritten in the Schr\"odinger picture and then extended to open quantum systems by relating onto each other two different evolutions of the $\mathcal{N}$-point functions of the open quantum system. As a concrete example of the mapping, we consider the mean-field dynamics of a simple dissipative quantum system that consists of a one-dimensional Bose-Einstein condensate being locally bombarded by a dissipating beam of electrons in both cases when the beam amplitude or the waist is steady and modulated.
%
\end{strip}

\section{Introduction}


The concrete and complete description of the dynamical behavior of any physical system can be given by correlations~\cite{Kukuljan2018}. For quantum systems with many degrees of freedom, the complete information can be obtained from quantum field theory using multitime-multipoint correlation functions. Using such functions, it is possible to directly express any observable of the system~\cite{WeinbergBook,Guarrera2011,Kukuljan2018}.
First-order (one-body) and second-order (two-body) correlations have been investigated for trapped cold gases by using involved interferometry in atom chip~\cite{Hellweg2003} or free expansion~\cite{Manz2010}. In quasi-one dimensional trapped cold gases, quantum fluctuations are enhanced, and first-order correlations are important in describing the coherence properties of the system, especially in the quasi-condensate regime when the system turns smoothly from a decoherent system to a finite-size Bose-Einstein condensate~\cite{Manz2010}.
Higher-order correlations and the way they decompose to lower orders are very important in many-body systems, since they provide useful information on the interactions, the structure and the complexity of the system~\cite{Schweigler2017,Schweigler1,Schweigler2} including, for instance, decoherence and thermalization~\cite{Giraud2010}, the discrimination between the ground and thermal states, and the spectrum of quasi-particles~\cite{Kukuljan2018}. Correlations of order one or two have been, however, the most frequently measured correlations. Recently, there has been immense progress in recipes for measuring higher-order correlations~\cite{Denis2017} to the ultimate goals of non-destructively exploring quantum systems~\cite{Guarrera2011,Hung2011} and achieving quantum-field tomography~\cite{Steffens2014}. In order to explore the quantum coherence properties of massive particles, it has been possible to observe higher-order correlations up to the sixth order~\cite{Dall2013}. The knowledge of correlations finds applications mostly in quantum metrology~\cite{Steffens2014,Steffens1}, quantum information~\cite{Steffens1,Steffens2}, and quantum simulations~\cite{Steffens3,Steffens4} where they allow reading, verifying and characterizing quantum systems. 

Relevant many-body quantum systems, both for closed and open cases, can be designed and controlled with an unprecedented precision due to recent technical development of quantum gas
experiments~\cite{Wurtz2009}. In open systems~\cite{Grigoriu2012,Koch2016,Shen2018}, in particular, localized dissipation provides new possibilities to engineer robust many-body quantum states and allows us to study fundamental quantum effects like quantum Zeno dynamics~\cite{BarontiniPRL2013,FroemlOtt2019} and non-equilibrium dynamics in an ultracold quantum gas~\cite{Labouvie2015,Labouvie2016}. Despite those achievements, however, numerous regimes of many-body quantum systems are still challenging both theoretically and experimentally.

 
 Motivated by those theoretical and experimental efforts for achieving and controlling open quantum systems, we here study the dynamics of a class of open quantum systems through an exact $\mathcal{N}$-point function mapping between different experimental situations. We aim at finding how the data of a pair of experiments can be mapped onto each other. One of the experiments may be interesting but challenging enough to be achieved, for instance, due to technical limitations. Our current contribution is restricted to equal-time multipoint correlation functions which does not capture autocorrelations, i.e., correlation functions for different times. But a lot of information can be obtained from quantum field theory using those simpler correlators. Moreover, using those correlators, for instance, one would not have to worry about the effects of wave function collapse from the first measurement changing the results of the second measurement.
%
The work is organized as follows. 
In Sec.~\ref{secA}, starting from previous results on the quantum-field mapping in the Heisenberg picture, we establish a corresponding result in the Schr\"odinger picture by expressing a pair of many-body wave functions in terms of second quantized fields, and then relating them onto each other.
 In Sec.~\ref{secA2}, 
 we present the spacetime mapping of a pair of $\mathcal{N}$-point functions that would represent two different experiments with lossy many-body quantum systems. The quantum-field mapping for closed systems formulated in Ref.~~\cite{WPA1} is rewritten in terms of wave functions, and then further generalized to open systems described by the Lindblad equation with loss or gain channels through the $\mathcal{N}$-point correlation functions. Such an equation has proven to be useful in describing the dynamics of Markovian open quantum systems, see for instance~\cite{BarontiniPRL2013,FroemlOtt2019}.
 Sec.~\ref{secB} deals with a concrete example that illustrates the mapping between two experimental situations where dissipation is created on a typical bosonic system by a Gaussian beam of electrons. Numerical computations are performed with the mean-field Gross-Pitaevskii equation to quantify our findings.
 The work ends in section~\ref{concl} which is devoted to the conclusion and summary of our results.


\section{Schr\"odinger-picture version of the exact quantum-field mapping for closed systems}  \label{secA}

Spacetime mappings have long been used, essentially as coordinate transformations to put theoretical problems into forms where they can easily be solved. A brief survey of solution-oriented mappings was provided in Ref.~\cite{WPA1}. Using spacetime transformations for mapping two different parameter regimes of an experiment is fundamental for physics. But to our knowledge this non-traditional use of mappings was formally proposed only recently as an exact quantum-field mapping for studying challenging dynamical regimes of isolated quantum gases~\cite{WPA1}. In what follows, taking advantage of such a mapping for closed systems~\cite{WPA1}, we formulate a corresponding Schr\"odinger-picture version of the mapping that is valid for both bosonic and fermionic systems.


\subsection{Revisiting the exact quantum-field mapping}

We consider a $D$-dimensional quantum gas which can be a mixture of several species of type $k$ with possibly different masses $M_k$ in any arbitrary state. The gas can be a single- or multi-component bosonic or fermionic system, or even a mixture of bosons and fermions. All observables of the gas may be expressed in terms of the time-evolving second-quantized field operator $\hat{\psi}_k(\mathbf{r},t)$, which destroys a particle, and of its Hermitian conjugate field operator $\hat{\psi}_l^\dagger(\mathbf{r},t)$, which correspondingly creates a particle of type $k$, at position $\mathbf{r}$ and time $t$. We require the field operators to satisfy the canonical (anti-)commutation relations
\begin{equation}
\left[\hat{\psi}_k(\mathbf{r},t), \hat{\psi}_l^\dagger(\mathbf{r}',t) \right]_{\pm} =\delta_{kl} \, \delta^D\left(\mathbf{r}-\mathbf{r}'\right),
\end{equation}
where $\left[\hat{A}, \hat{B} \right]_{\pm} = \hat{A} \hat{B} \pm \hat{B} \hat{A} $, such that the theory can be equally applicable to fermions (anticommutation) and bosons (commutation)~\cite{WPA1}. There has been tremendous experimental progress in creating and monitoring various kinds of dynamics of ultracold quantum gases with a large range of two-particle interaction, $U_{klmn}(\mathbf{r},\mathbf{r}',t)$, and one-particle interaction, $V_k(\mathbf{r},t)$, imposed by external trapping fields. Without loss of generality, we will drop the species type subscript index in what follows. In general, experimental measurements in a quantum gas can be expressed in terms of an $\mathcal{N}$-point correlation function. The $\mathcal{N}$-point or $\mathcal{N}$-body correlation function is a general expectation value of the field operators that describe the quantum gas properties. As from here, we omit the species indices to simplify the notations. Our theory is general, but we restrict ourselves to spatial correlations, and thus the equal time $\mathcal{N}$-point function can be written as~\cite{WPA1,Naraschewski1999,Dall2013}:
\begin{eqnarray} \label{Npoint}
\mathcal{F}\left(\mathbf{R},\mathbf{R}',t\right) = \left\langle \left[\prod_{j=1}^\mathcal{N} \hat{\psi}^\dagger\left(\mathbf{r}_{j'} ', t\right) \right]  \left[\prod_{j=1}^\mathcal{N} \hat{\psi}\left(\mathbf{r}_j, t\right) \right] \right\rangle,
\end{eqnarray}
where $\mathbf{R}=\left(\mathbf{r}_1, ..., \mathbf{r}_\mathcal{N}\right)$ and $\mathbf{R}'=\left(\mathbf{r}_1', ..., \mathbf{r}_\mathcal{N}' \right)$. For notational simplicity, we introduced the subscript index $j'\equiv \mathcal{N}+1-j$. It is worth noting that the order of operators in the $\mathcal{N}$-point function is quite rigid for fermions due to the anticommutativity of annihilation/creation operators. For bosons, setting $j'=j$ would still be a correct notation. Let us mention in passing that the $\mathcal{N}$-point function $\mathcal{F}\left(\mathbf{R},\mathbf{R}',t\right)$ is sometimes denoted $G^{(\mathcal{N})}\left(\mathbf{R},\mathbf{R}'\right)$.

We consider a quantum gas that evolves under two different experimental conditions, but in a related way, for some particular pairs of interparticle interactions $(U,\tilde{U})$ and trapping potentials $(V,\tilde{V})$. Suppose both evolutions are described by a pair of different sets of quantum fields $(\hat{\psi},\hat{\Psi})$, and their conjugates $(\hat{\psi}^\dagger,\hat{\Psi}^\dagger)$. The evolutions with $\{\hat{\psi},V,U\}$ and $\{\hat{\Psi},\tilde{V},\tilde{U}\}$ thus satisfy the same Heisenberg equation as well as canonical (anti-)commutation relations. The Heisenberg-picture quantum-field operators are mapped as follows~\cite{WPA1}:
\begin{eqnarray}\label{H.mapping}
\hat{\Psi}(\mathbf{r},t) = e^{-\frac{i M}{2\hbar}\frac{\dot{\lambda}}{\lambda}r^{2}}\lambda^{D/2}\hat{\psi}\left(\lambda\mathbf{r},\tau\right) ,
\end{eqnarray}
where $\lambda = \lambda(t)$ is the free parameter of the problem. It fullfills the two conditions $\lambda(0)=1$ and $\dot{\lambda}(0)=0$, where $\dot{\lambda}(t)\equiv d\lambda/dt$ and $d\tau/dt = \lambda^{2}$ with $\tau=\tau(t)$, so that the fields in both experiments coincide at initial time, meaning that the corresponding experiments are supposed to be run with almost identical initial state. This can be achieved practically by preparing a state, and then adiabatically splitting it into two. For an interparticle interaction with homogeneity degree $s$ (equal to $D$ for contact interaction), the potentials map as
\begin{eqnarray}\label{H.mapping1}
\tilde{U}(\mathbf{r},\mathbf{r}',t) = \lambda^{2} U(\lambda\mathbf{r},\lambda\mathbf{r}',\tau(t)) = \lambda^{2-s} U(\mathbf{r},\mathbf{r}',t) , \nonumber \\
\tilde{V}(\mathbf{r},t) = \lambda^{2} \left[ V(\lambda\mathbf{r},\tau(t)) + \frac{1}{2}M r^2 \lambda \hat{\Omega}^2 \lambda \right],
\end{eqnarray}
where we use the time differential operator $\hat{\Omega}=\lambda^{-2}\partial/\partial t \equiv \partial/\partial\tau$.
Using the quantum-field mapping above, we obtain that the quantum gas experiment described by the $\mathcal{N}$-point function~\eqref{Npoint} can be mimicked by an equivalent experiment described by a rescaled $\mathcal{N}$-point function $\tilde{\mathcal{F}}$, such that
\begin{align}\label{Npoint1}
\begin{split}
\tilde{\mathcal{F}}\left(\mathbf{R},\mathbf{R}',t\right) = \left\langle \left[\prod_{j=1}^\mathcal{N} \hat{\Psi}^\dagger\left(\mathbf{r}_{j'} ', t\right) \right]  \left[\prod_{j=1}^\mathcal{N} \hat{\Psi}\left(\mathbf{r}_j, t\right) \right] \right\rangle .
\end{split}
\end{align}
In this expression, we used again $j'\equiv \mathcal{N}+1-j$. We find that the $\mathcal{N}$-point functions $\tilde{\mathcal{F}}$ and $\mathcal{F}$ of the two experimental situations are related onto each other through the mapping identity
\begin{eqnarray}\label{H.mapping2}
\tilde{\mathcal{F}}\left(\mathbf{R},\mathbf{R}',t\right) =  \lambda^{\mathcal{N} D} e^{-\frac{i M}{2\hbar}\frac{\dot \lambda}{\lambda}\sum \limits_{i=1}^{\mathcal{N}}(\mathbf{r}_i^2-\mathbf{r}_i'^2)} \mathcal{F}\left(\lambda \mathbf{R},\lambda \mathbf{R}',\tau \right).
\end{eqnarray}
Here, the scaling factor is not only a function of the dimension $D$, but also a function of the number $\mathcal{N}$.

\subsection{Expressing the wave function in terms of second-quantized fields}

The dynamics of a quantum system can be mathematically formulated in the two equivalent and most important  representations of quantum mechanics, which are the Schr\"odinger and Heisenberg pictures. A natural question that emerges is how the exact mapping reads in the Schr\"odinger picture, which is the originally approved and most popular representation. Unlike the Heisenberg picture, the Schr\"odinger picture regards the quantum-field operators associated to all physical quantities of a quantum-mechanical system to be fundamentally fixed in time. Meanwhile quantum systems are represented by time evolving wave functions or state kets.
The Schr\"odinger-picture quantum fields can readily be expressed in terms of Heisenberg-picture quantum fields by introducing a unitary operation $\hat{\mathcal{U}}(t,t')$ such that $\hat{\psi}_\mathrm{H} = \hat{\mathcal{U}}^\dagger \hat{\psi}_\mathrm{S} \, \hat{\mathcal{U}}$.
The subscripts $S$ and $H$ refer to Schr\"odinger and Heisenberg representations, respectively. In what follows, we will omit the subscripts for notational simplicity, but Heisenberg variables will still be identifiable through an explicit time $t$ in the arguments or subscripts. Thus we can write
\begin{eqnarray}\label{unitary}
\hat{\psi}(\mathbf{r},t) = \hat{\mathcal{U}}^\dagger \hat{\psi}(\mathbf{r}) \, \hat{\mathcal{U}}, ~~ \ket{\psi_t} = \hat{\mathcal{U}}\ket{ \psi }. 
\end{eqnarray}
For the sake of clarity, we recall the following textbook formulations which are equivalent: 
For the Heisenberg-picture field operator at time $t$, we have $\hat{\psi}(\mathbf{r},t)=\hat{\psi}_t(\mathbf{r}) \equiv \hat{\psi}_\mathrm{H}$. For the Schr\"odinger-picture field operator at all times (corresponding to the Heisenberg-picture field operator at time $t=0$), we have $\hat{\psi}(\mathbf{r})=\hat{\psi}(\mathbf{r},0) =\hat{\psi}_0(\mathbf{r},0) \equiv \hat{\psi}_\mathrm{S}$. For the Schr\"odinger-picture state at time $t$, we have $\ket{\psi_t}=\ket{\psi}_\mathrm{S}$. And for the Heisenberg-picture state at all times (corresponding to Schr\"odinger-picture state at time $t=0$), we have $\ket{\psi_0}=\ket{\psi} \equiv \ket{\psi}_\mathrm{H}$.


Equation~\eqref{unitary} above relates both second-quantized field operators $\hat{\psi}(\mathbf{r})$ and $\hat{\psi}(\mathbf{r},t)$ in Schr\"odinger and Heisenberg pictures, respectively, providing a way of writing the wave function and the $\mathcal{N}$-point function~\eqref{Npoint} using non time-evolving field operators.
Let $\ket{0}$ be the (position-space) vacuum state ket, $\ket{\psi_t}$ the state ket of the system at time $t$, and $\psi \left(\mathbf{r}_1, \cdots , \mathbf{r}_N,t\right) $ the position-space wave function for the gas with $N$ particles, where $\mathbf{r}_j$ is the position of the $j$th particle in $D$-dimensional space. Then the wave function can be obtained by projecting the state ket on the position space:
\begin{eqnarray} 
\psi \left(\mathbf{r}_1, \cdots , \mathbf{r}_N,t\right) 
& \equiv &  \langle \mathbf{r}_1, \cdots , \mathbf{r}_N | \psi_t \rangle \nonumber \\
 &=& \Big\langle 0 \Big| \prod  \limits_{j=1}^N \hat{\psi}(\mathbf{r}_j) \Big| \psi_t \Big\rangle . \label{Wavefunction1}
\end{eqnarray}
The many-body wave function $\psi$ describes all the particles at a time and satisfies the following normalization condition
\begin{eqnarray}
 \int  \Big(\prod_{j=1}^N d^{D}\mathbf{r}_j \Big)   \: \psi ^* \psi = N.
\end{eqnarray}
Unlike for second-quantized field operators, the number of particles $N$ explicitly appears in the many-body wave function as well as in the many-body Hamiltonian, see below.


Using the unitary transformation~\eqref{unitary}, we can transfer the time dependence from the state to the field operator and rewrite the many-body wave function~\eqref{Wavefunction1} into the following form 
\begin{eqnarray}  \label{Wavefunction2}
\psi \left(\mathbf{r}_1, \cdots , \mathbf{r}_N,t\right) 
= \Big\langle 0 \Big| \prod  \limits_{j=1}^N \hat{\psi}(\mathbf{r}_j,t) \Big| \psi \Big\rangle .
\end{eqnarray}
The many-body wave function is expressed in terms of evolving quantum fields. This opens a possibility for using the quantum-field mapping to construct the mapping of many-body wave functions.

\subsection{Mapping identity for many-body wave functions}


Consider two potential experiments A and B performed with the same Heisenberg (stationary) state and different Hamiltonians. Each of both evolutions with $\{\psi,V,U\}$ and $\{\Psi,\tilde{V},\tilde{U}\}$ satisfies the following Schr\"odinger equation:
\begin{eqnarray} 
i \hbar \frac{\partial }{\partial t} {  \ket{\psi_t}}  = H \ket{\psi_t},
\end{eqnarray}
where the Schr\"odinger Hamiltonian for the $D$-dimensional quantum gas with $N$ particles can be written as
\begin{eqnarray} 
H =  \sum \limits _{j=1}^N \Big(- \frac{\hbar^2}{2M_j} \nabla_{\mathbf{r}_j}^2 +V(\mathbf{r}_j,t)\Big) \nonumber \\
 +  \sum_{j=1}^N \sum_{k>j, k=1}^N  U(\mathbf{r}_k,\mathbf{r}_j,t).
\end{eqnarray} \label{Hamil2}
The pairs of potentials $(V,U)$ and $(\tilde{V},\tilde{U})$ can be related onto each other according to Eq.~\eqref{H.mapping1}. It should be noted that the time $t$ in the Hamiltonian represents an explicit time-dependence, and not the result of an evolution of the Hamiltonian operator under a unitary transformation.


As for the wave function, it is a key tool of the Schr\"odinger picture and thus basically defined in terms of second quantized fields in Schr\"odinger picture. Using Eq.~\eqref{Wavefunction2} the many-body wave functions at time $t$  corresponding to the two experiments B and A, however, can be written as functions of second quantized fields. We get
\begin{eqnarray}  \label{Wavefunction3}
\Psi \left(\mathbf{r}_1, \cdots , \mathbf{r}_N,t\right) 
= \Big\langle 0 \Big| \prod  \limits_{j=1}^N \hat{\Psi}(\mathbf{r}_j,t) \Big| \Psi \Big\rangle , 
\nonumber \\
\psi \left(\lambda \mathbf{r}_1, \cdots , \lambda \mathbf{r}_N,\tau\right) 
= \Big\langle 0 \Big| \prod  \limits_{j=1}^N \hat{\psi}(\lambda \mathbf{r}_j,\tau) \Big| \psi \Big\rangle .
\end{eqnarray}
%
%
%
Since the exact space-time mapping is known for evolving quantum-field operators, it becomes clear that both wave functions can be mapped onto each other. That can be done by combining together Eqs.~\eqref{H.mapping} and \eqref{Wavefunction3}. We obtain that the position-space wave functions in the two potential experiments will be exactly related through the following identity
\begin{eqnarray} 
\Psi \left(\mathbf{r}_1, \cdots , \mathbf{r}_N;t\right)   
=   &\lambda^{\frac{N D}{2}} & e^{-i \frac{M}{2\hbar} \frac{\dot{\lambda}}{\lambda}  \sum\limits_{j=1}^N \mathbf{r}_j^2} \nonumber \\ 
& \times &
\psi \left(\lambda\mathbf{r}_1, \cdots , \lambda\mathbf{r}_N;\tau(t) \right) .
\end{eqnarray}
Similarly to the quantum-field mapping, the exact wave function mapping consists of a non trivial time-dependent transformation of space and a multiplication of many-body wave functions by a Gaussian phase factor. Here that factor explicitly depends on the number of particles and their positions at a given time.

\section{Mapping of different evolutions of $\mathcal{N}$-point functions in an open system}\label{secA2}


The result above is obtained for closed quantum systems. Here we aim at generalizing the identities to dissipative systems described by the Lindblad master equation by mapping the $\mathcal{N}$-point functions of their evolutions.
We start by presenting the useful ingredients in Schr\"odinger picture. Notably, we rewrite the $\mathcal{N}$-point function using second quantized fields in the Schr\"odinger picture and present the appropriate Lindblad equation. Afterwards we derive the Schr\"odinger-picture evolution of an $\mathcal{N}$-point function of a quantum gas that obeys the Lindblad equation. Then invoking the Heisenberg evolution and the picture-independence of expectation values, which makes equivalent all pictures of quantum mechanics, we present a mapping identity for relating onto each other two different evolutions of $\mathcal{N}$-point functions. Note that the corresponding calculations are performed for both bosonic and fermionic systems.

\subsection{Necessary tools in Schr\"odinger picture}

Obtaining the mapping of $\mathcal{N}$-point functions requires the use of second quantized fields in the Schr\"odinger picture into the appropriate Lindblad equation. Those ingredients are presented in this section.

\subsubsection{The $\mathcal{N}$-point function in terms of second quantized fields in the Schr\"odinger picture}

In order to get the $\mathcal{N}$-point function in terms of Schr\"odinger-picture quantum-field operators, we can start from the definition of $\mathcal{N}$-point functions~\eqref{Npoint}, and then apply the transformation~\eqref{unitary} and its conjugate. Since $\mathcal{N}$-point functions are expectation values, they remain the same irrespective of the quantum-mechanical representation used to describe them. For that, we can write
\begin{align} \label{Npoint2}
\begin{split}
\mathcal{F}\left(\mathbf{R},\mathbf{R}',t\right) = \left\langle \left[\prod_{j=1}^\mathcal{N} \hat{\psi}^\dagger\left(\mathbf{r}_{j'} '\right) \right]  \left[\prod_{j=1}^\mathcal{N} \hat{\psi}\left(\mathbf{r}_j\right) \right] \right\rangle ,
\end{split}
\end{align}
where $j'\equiv \mathcal{N}+1-j$, and $\mathcal{F}$ now denotes the $\mathcal{N}$-point function in the Schr\"odinger picture. In contrast to~\eqref{Npoint}, the time dependence in the equation above is actually only implicit. It is effectively redeemed, however, when the non-evolving operators are applied onto evolving state kets or wave functions. In both the Heisenberg and Schr\"odinger pictures, we have kept the same notation for the $\mathcal{N}$-point function. 

In the Schr\"odinger picture, since operators associated to physical quantities are non time evolving, the evolution of observables is given by the equations of motion for the state kets onto which operators can be applied. The time dependence of all observables is thus determined by the time dependence of the state kets or of the density matrix, and a master equation in the Lindblad form may describe the dynamics of the dissipative quantum system.

\subsubsection{The Lindblad master equation}

Quantum gases are in general well isolated from the environment or bath. However, particle losses and heating can occur, for instance, due to background gas collisions, and photon scattering and trap shaking, respectively.
We restrict ourselves to the case of systems that satisfy the Markovian (or Born-Markov) approximation. Then the connection between the system and the bath is so weak that the bath forgets any information received from the system much faster than the  evolution  we want  to follow.  We then expect a clean  separation  between  the  typical  correlation  time  of  the system's fluctuations  and  the  time  scale  of  the  evolution described.
In that case, the evolution of the density matrix and the full quantum fields of the system should be based on a master equation in Lindblad form~\cite{Lindblad1976}. The Lindblad formulation is interesting because not only it is less difficult to solve compared to non-Markovian equations, but also it may allow describing the behavior of both bosonic~\cite{Prosen2008,Braaten2017} and fermionic~\cite{Kordas2015} systems. Moreover, it turns out to give an acceptably accurate description of recent experimental efforts to control dissipative quantum systems of bosons~\cite{BarontiniPRL2013} or fermions~\cite{FroemlOtt2019}, see~\cite{Weiss2008Book} for an overview. We use such a master equation as the theoretical model governing the dynamics of the class of quantum systems studied in this paper. In its standard form in the Schr\"odinger picture, the Lindblad equation reads~\cite{Breuer2002Book}:
\begin{align}\label{Lindblad}
\begin{split}
&i\hbar \, \frac{\partial  \hat\rho}{\partial t}  = \left[\hat H ,   \hat\rho  \right]_{-} + i \hbar \, \mathcal{\hat L}[\hat\rho],
\end{split}
\end{align}
where $\hat H$ in the von Neumann term is the Hamiltonian of the system expressed in terms of second quantized fields in the Schr\"odinger picture as 
\begin{align}\label{Hamil}
\begin{split}
\hat H = \int d^D \mathbf{r}\,  \hat{\psi}^{\dagger}(\mathbf{r})\left( -\frac{\hbar^2}{2M}\nabla^2 + V(\mathbf{r},t) \right)  \hat{\psi}(\mathbf{r}) \\
+\frac{1}{2}  \int d^D \mathbf{r} d^D \mathbf{r}'\,  \hat{\psi}^{\dagger}(\mathbf{r}) \hat{\psi}^{\dagger}(\mathbf{r}')U\left(\mathbf{r},\mathbf{r}',t \right)  \hat{\psi}(\mathbf{r}') \hat{\psi}(\mathbf{r}) .
\end{split}
\end{align}
Even though Eq.~\eqref{Lindblad} cannot describe general open systems since it is an approximate quantum master equation, it suitably describes the outcomes of quantum gas experiments as pointed out above. As it can be seen, the quantum fields operators are not time-dependent, only the potentials have an explicit time-dependence, \textit{i.e.}, their time change is not due to the action of a quantum mechanical operator. The Lindbladian term for lossy systems may read
\begin{align} 
\mathcal{L}[\hat{\rho}] 
&=  \sum_{j=1}^{\mathcal{D}^2-1} \gamma_{j} \left( \hat{Q}_{j}  \hat{\rho} \hat{Q}_{j}^\dagger - \frac{1}{2}\left[ \hat{\rho} , \hat{Q}_{j}^\dagger  \hat{Q}_{j}   \right]_{+}  \right).
\end{align}
$\mathcal{D}$ denotes the dimension of the system's Hilbert space and ${\rho}$ represents the density matrix of the reduced (open) system. Furthermore, $\hat{Q}_j$ are Lindblad operators (or quantum jump operators) which represent arbitrary linear operators taken in the Hilbert space of the system, each acting as an independent incoherent process.
Those Lindblad generators may be taken to be time-dependent, which allows a more general description of open systems that encloses the systems subjected to an external time-dependent field. 
In the case of a one-dimensional leaky lattice, for instance, the correlated two- and three-body particle losses are expressed in terms of dissipation~\cite{Trimborn2011,Witthaut2011,Dast2014,Torres2014} 
and $\gamma_j$ denote the respective loss or coupling rates at sites $j$. Meanwhile $\hat{Q}_j^\dagger$ and $\hat{Q}_j$ play the roles of the particle creation and annihilation operators for bosons, respectively.
Apart from losses, the Lindblad equation may also include gains in some terms of the sum. The balanced loss and gain can strongly affect the properties of many-body systems. It has been shown that the balanced loss and gain between lattice sites lead to the existence of stationary states and the phase shift of pulses between two lattice sites in many-particle systems~\cite{Dast2014}. In bosonic many-body quantum systems, dissipation can interplay with interparticle interactions, which affects the coherence and the density fluctuations of the system~\cite{Poletti2012}.

We consider a more general system where dissipation is continuous and time-dependent but with only a single kind of dissipative process. The Lindbladian may be written as
\begin{align}
\begin{split}
& \mathcal{\hat L}[\hat{\rho}] =-\int d^D \mathbf{r}\,   \left( \hat{Q}^{\dagger}\hat{Q} \hat\rho  +\hat\rho \hat{Q}^{\dagger}\hat{Q} - 2\hat{Q} \hat\rho\hat{Q}^{\dagger} \right),
\end{split}
\end{align}
where the jump operators can be space- and time-dependent. Their specific form depends on the kind of coupling between the system and its environment.
Choosing the jump operators to feature a pure loss, \textit{i.e.}, $\hat{Q}=\hat{\psi}(\mathbf{r})\sqrt{\gamma(\mathbf{r},t)/2}$, we obtain
\begin{align}\label{Lindbladian}
\begin{split}
& \mathcal{\hat L}[\hat{\rho}] =-\int d^D \mathbf{r}\, \frac{\gamma(\mathbf{r},t)}{2} \left( \hat\psi^{\dagger}\hat\psi \hat\rho  +\hat\rho \hat\psi^{\dagger}\hat\psi - 2\hat{\psi} \hat\rho\psi^{\dagger} \right).
\end{split}
\end{align}
 The jump operator used in this Lindbladian is a general annihilation operator applicable to both bosons and fermions. The dynamics of a system in a regime where the only dissipation channel is a gain, can readily be reproduced from the current work by taking the jump operator to be $\hat{Q}=\hat{\psi}^\dagger(\mathbf{r})\sqrt{\gamma_+(\mathbf{r},t)/2}$ where $\gamma_+$ is the gain rate. A regime with both gain and loss channels can also be readily deduced by simply summing the corresponding Lindbladians to get the full Lindbladian.

The standard Lindblad master equation~\eqref{Lindblad}, which is relevant to the Schr\"odinger picture, is expressed in terms of the non evolving second-quantized field operators. The Lindblad evolution based on time evolving second-quantized field operators, which are convenient for the Heisenberg picture, will be discussed below.

\subsection{Lindblad evolution of $\mathcal{N}$-point functions}\label{LindSec}

The time dependence of expectation values is completely carried out by the density matrix in the Schr\"odinger picture. The equation of motion of the expectation value of any arbitrary operator $\hat{\mathcal{O}}$ is given by the evolution rule
\begin{equation}\label{rule}
\frac{\partial \langle\hat{\mathcal{O}}\rangle }{\partial t} = \mathrm{Tr}\left( \hat{\mathcal{O}} \frac{\partial \hat{\rho}}{\partial t} \right) .
\end{equation}
Note that the explicit time dependence of Schr\"odinger's field operators is here ignored. As one can readily see, the $\mathcal{N}$-point function is expressed in the form $\langle\hat{\mathcal{O}}\rangle \equiv \mathcal{F}$, with $\hat{\mathcal{O}}$ being defined in terms of the set of operators $\{\hat\psi_{i}\}$ and their adjoints at different points. Thus we can write
\begin{equation}\label{EqA2}
\begin{split}
\frac{\partial \mathcal{F} }{\partial t}  
& =    \frac{1}{i \hbar} \left\langle \left[\hat{\mathcal{O}} , \hat{H} \right]_{-}  \right\rangle  + \mathrm{Tr}\left( \hat{\mathcal{O}}\mathcal{\hat{L}}\hat{\rho} \right).
\end{split}
\end{equation}
Using the Schr\"odinger-picture form of the $\mathcal{N}$-point function~\eqref{Npoint2} and the Lindblad equation~\eqref{Lindblad} into the rule~\eqref{EqA2}, exploiting the mathematical properties of the trace operator leads to the equation sought for. Let us provide the detailed derivation of the equation.


%
For simplification purpose, let us rewrite the $\mathcal{N}$-point function in terms of a general operator 
\begin{equation}\label{op}
\hat{\mathcal{O}}=\hat\psi^{'\dagger}_{\mathcal{N}}...\hat\psi^{'\dagger}_1 \hat\psi_1...\hat\psi_{\mathcal{N}}
= \left(\prod_{j=1}^{\mathcal{N}} \hat\psi^{'\dagger}_{j'}\right) \prod_{j=1}^{\mathcal{N}} \hat\psi_{j} .
\end{equation}
Notice that the index $j'\equiv \mathcal{N}+1-j$ is used. $\mathcal{F}$ in this case is a single time multipoint correlation function. The outline of the derivation of the time evolution equation of the $\mathcal{N}$-point function can be given as follows. We expand separately the two terms in the right-hand side of Eq.~\eqref{EqA2}, considering the hermiticity of the Hamiltonian, i.e., $\hat H^\dagger=\hat H$.

\subsubsection{Dissipation term}	


From Eq.~\eqref{Lindbladian}, we can write the dissipation term of Eq.\eqref{EqA2} as
\begin{align} 
\begin{split}
&{\textrm Tr}\left(\hat{\mathcal{O}} \mathcal{\hat L}\hat\rho \right) 
= -\int d^{D}\mathbf{r}''\, \frac{\gamma(\mathbf{r}'',t)}{2} \\
& \times   {\textrm Tr}\left( \hat{\mathcal{O}} \hat\psi^{''\dagger}\hat\psi'' \hat\rho +\hat{\mathcal{O}} \hat\rho \hat\psi^{''\dagger}\hat\psi''     - 2\hat{\mathcal{O}} \hat\psi''\hat\rho \hat\psi^{''\dagger} \right).
\end{split}
\end{align}
Using the cyclicity (invariance under cyclic permutation) and additivity properties of the trace operator, we express the dissipation term in terms of two useful commutators as follows: 
\begin{align} \label{trace1}
\begin{split}
{\textrm Tr}\left(\hat{\mathcal{O}} \mathcal{\hat L}\hat\rho \right) 
&=-\int d^{D}\mathbf{r}''\, \frac{\gamma(\mathbf{r}'',t)}{2}   {\textrm Tr}\left( \hat\psi^{''\dagger} [\hat\psi'', \hat{\mathcal{O}} ]_{-}\hat\rho\right) 
\\
& +\int d^{D}\mathbf{r}''\, \frac{\gamma(\mathbf{r}'',t)}{2}   {\textrm Tr}\left( [\hat\psi^{''\dagger},\hat{\mathcal{O}}]_{-}\hat\psi'' \hat\rho \right) .
\end{split}
\end{align}


Remember that all possible pairwise combinations of the $\mathcal{N}$ annihilation operators in the set $\{\hat\psi_j\}_{j=1}^\mathcal{N}$ commute (anticommute) for bosons (fermions), which is also the case for the set of corresponding creation operators $\{\hat\psi_j^{' \dagger}\}_{j=1}^\mathcal{N}$. In the following derivation we deal with bosons and fermions simultaneously by using in a smart way the following (anti)commutator identity
\begin{align} \label{ABC}
\begin{split}
[\hat{A}, \hat{B}\hat{C}]_{-}   =[\hat{A}, \hat{B}]_{\pm}\hat{C} \mp  \hat{B}[\hat{A},\hat{C}]_{\pm} ,
\end{split}
\end{align}
where $+$ denotes the anticommutator and $-$ the commutator.
Then using the canonical commutation and anticommutation rules for bosons and fermions, respectively, we can evaluate the two key commutators in the right-hand side of Eq.~\eqref{trace1}. We obtain
\begin{align} 
\begin{split}\label{comm}
&[\hat\psi^{''},\hat{\mathcal{O}}]  =\sum_{j=1}^{\mathcal{N}}\left[(-1)^{j''} \delta^D\left(\mathbf{r}{''}-\mathbf{r}_j^{'}\right) \frac{\partial \hat{\mathcal{O}}}{\partial \hat\psi_j^{'\dagger}}   \right] , 
\\ 
&[\hat\psi^{''\dagger},\hat{\mathcal{O}}]  = \sum_{j=1}^{\mathcal{N}}\left[ (-1)^{1+j''}\delta^D\left(\mathbf{r}{''}-\mathbf{r}_j\right) \frac{\partial \hat{\mathcal{O}}}{\partial \hat\psi_j}   \right].
\end{split}
\end{align}
We have $j''=0$ and $j''=\mathcal{N}-j$ for bosons and fermions, respectively. The factor with $-1$ is reminiscent of anticommutativity of fermionic fields since a negative sign would appear each time two annihilation/creation operators are swapped in the general operator $\hat{\mathcal{O}}$ for a fermion. As a consequence, we have
\begin{align} 
\begin{split}\label{deriv}
&\hat\psi^{'\dagger}_j  \frac{\partial \hat{\mathcal{O}}}{\partial \hat\psi_j^{'\dagger}}  =  \frac{\partial \hat{\mathcal{O}}}{\partial \hat\psi_j}   \hat\psi_j  =  (-1)^{j''}\hat{\mathcal{O}} .
\end{split}
\end{align}
A subsequent substitution of Eqs.~\eqref{comm} and \eqref{deriv} into Eq.~\eqref{trace1} yields 
\begin{strip}
	\begin{align} 
	\begin{split}
	{\textrm Tr}\left(\hat{\mathcal{O}} \mathcal{\hat L}\hat\rho \right)  = & \frac{1}{2}\sum_{j=1}^{\mathcal{N}} (-1)^{1+j''} \left[ \gamma(\mathbf{r}'_j,t)  {\textrm Tr} 
	\left(\hat\psi^{'\dagger}_j  \frac{\partial \hat{\mathcal{O}}}{\partial \hat\psi_j^{'\dagger}} \hat\rho \right) + \gamma(\mathbf{r}_j,t)   {\textrm Tr} 
	\left(\frac{\partial \hat{\mathcal{O}}}{\partial \hat\psi_j}   \hat\psi_j \hat\rho \right) \right]
	= -\sum_{j=1}^{\mathcal{N}} \frac{\gamma(\mathbf{r}'_j,t)}{2}   
	\left\langle \hat{\mathcal{O}}  \right\rangle -\sum_{j=1}^{\mathcal{N}} \frac{\gamma(\mathbf{r}_j,t)}{2}   
	\left\langle \hat{\mathcal{O}}  \right\rangle.
	\end{split}
	\end{align}
\end{strip}
Thus the dissipation term for the $\mathcal{N}$-point function yields

\begin{align} \label{Dissip}
\begin{split}
i\hbar {\textrm Tr}\left(\hat{\mathcal{O}} \mathcal{\hat L}\hat\rho \right) 
=&
- i \hbar \, \sum_{j=1}^{\mathcal{N}} \frac{\gamma(\mathbf{r}_j,t)+\gamma(\mathbf{r}'_j,t)}{2}   
\left\langle \hat{\mathcal{O}}  \right\rangle .
\end{split}
\end{align}

\subsubsection{The von Neumann term}

In order to expand the Hamiltonian term of Eq.\eqref{EqA2}, let $\hat A_n=\hat\psi^{'\dagger}_{n}...\hat\psi^{'\dagger}_1 \hat\psi_1...\hat\psi_{n}$, with $n=1,...,\mathcal{N}$, which means $\hat A_\mathcal{N}=\hat{\mathcal{O}}$. Then we can express the commutator $\left[ \hat A_{n}, \hat H   \right]_{-}$ in terms of next commutator $\left[ \hat A_{n-1}, \hat H   \right]_{-}$ as
\begin{align} \label{formul}
\begin{split}
\left[ \hat A_{n}, \hat H   \right]_{-}  
=&  \left[ \hat\psi^{'\dagger}_{n} \hat A_{n-1} \hat\psi_{n}, \hat H   \right]_{-}   \\
=&
\hat\psi^{'\dagger}_{n} \left[\hat A_{n-1} , \hat H   \right]_{-}\hat\psi_{n}
\\&+\hat\psi^{'\dagger}_{n} \left(-\overleftarrow{h}_{n}^{'} \hat A_{n-1}+\hat A_{n-1}  \overrightarrow{h}_{n} \right) \hat\psi_{n} .
\end{split}
\end{align}
 The energy operators $\overrightarrow{h}_{j} \equiv h(\mathbf{r}_{j}) $ and $\overleftarrow{h}_{j}^{'}  \equiv h(\mathbf{r}_{j}')$ apply only rightwise on $\hat\psi_{j}$ and leftwise on $\hat\psi^{'\dagger}_{j}$, respectively. The function $h(\mathbf{r}_{j})$ is given by the expression 
\begin{align}\label{ham0}
\begin{split}
h(\mathbf{r})=  -& \frac{\hbar^2}{2M}  \nabla_{\mathbf{r}}^2  +  V (\mathbf{r},t)   	\\
& +
\int d^{D}\mathbf{r}''  \,  \hat{\psi}^{\dagger} (\mathbf{r}'') U(\mathbf{r}'',\mathbf{r},t)     {\hat{\psi}} (\mathbf{r}'') ,
\end{split}
\end{align}
which is nothing but a second-quantized many-particle Hamiltonian. Eq.~\eqref{NpointEvol1} provides the evolution of $\mathcal{N}$-point functions in a dissipative quantum system described by the Lindblad equation.
Let us iterate the formula~\eqref{formul} down to $n=1$. For illustration, the first two iterates are
\begin{align} 
\begin{split}
\left[ \hat A_{n}, \hat H   \right]_{-}  
=&
\hat\psi^{'\dagger}_{n} \left[\hat A_{n-1} , \hat H   \right]_{-} \hat\psi_{n}
\\+&\hat\psi^{'\dagger}_{n} \left(-\overleftarrow{h}_{n}^{'} \hat A_{n-1}+\hat A_{n-1}  \overrightarrow{h}_{n} \right) \hat\psi_{n} \\
\left[ \hat A_{n-1}, \hat H   \right]_{-}  
=&
\hat\psi^{'\dagger}_{n-1} \left[\hat A_{n-2} , \hat H   \right]_{-} \hat\psi_{n-1}
\\+&\hat\psi^{'\dagger}_{n-1} \left(-\overleftarrow{h}_{n-1}^{'} \hat A_{n-2}+\hat A_{n-2}  \overrightarrow{h}_{n-1} \right) \hat\psi_{n-1},
\end{split}
\end{align}
and the last two iterates are
\begin{align} 
\begin{split}
\left[ \hat A_{2}, \hat H   \right]_{-} 
=&
\hat\psi^{'\dagger}_{2} \left[\hat A_{1} , \hat H   \right]_{-} \hat\psi_{2} \\
+& \hat\psi^{'\dagger}_{2} \left(-\overleftarrow{h}_{2}^{'} \hat A_{1}+\hat A_{1}  \overrightarrow{h}_{2} \right) \hat\psi_{2}\\
\left[ \hat A_{1}, \hat H   \right]_{-} 
=& 
\hat\psi^{'\dagger}_{1} \left(-\overleftarrow{h}_{1}^{'} +  \overrightarrow{h}_{1} \right) \hat\psi_{1} .
\end{split}
\end{align}
We introduce the quantity $\hat A_0=\mathbb{I}$ (identity). Using backward substitution in the above set of equations up to $n=\mathcal{N}$ we obtain the following expression for the commutator
\begin{align} \label{Hamil3}
\begin{split}
\left[ \hat{\mathcal{O}}, \hat H   \right]_{-}  
= \sum_{j=0}^{\mathcal{N}-1}\Bigg[ \left(\prod_{i=1}^{\mathcal{N}-j}\hat\psi^{'\dagger}_{\mathcal{N}+1-i}\right)  \Big( \hat A_{j} \overrightarrow{h}_{j+1}\\-\overleftarrow{h}_{j+1}^{'} \hat A_{j}\Big)  
\times
\left(\prod_{i=j+1}^{\mathcal{N}}\hat\psi_{i}\right)  \Bigg].
\end{split}
\end{align}
For the validity of the above equation, the following conventions are assumed:
\begin{equation}
\hat{A}_0=\mathbb{I}, \: \prod_{i=1}^{0} \hat\Psi_{\mathcal{N}+1-i}^{'\dagger}=\mathbb{I}, \text{ and } \prod_{i=\mathcal{N}+1}^{\mathcal{N}} \hat\Psi_i=\mathbb{I} .
\end{equation}Let us mention in passing that we can alternatively derive that relation through a substitution of Eqs.~\eqref{Hamil}, \eqref{ABC} and \eqref{comm} into the commutator $\left[ \hat{\mathcal{O}}, \hat H   \right]_{-}$.

\begin{strip}
	
\subsubsection{The evolution equation}

Putting the dissipation and Hamiltonian terms \eqref{Dissip} and \eqref{Hamil3} together, we finally get the evolution equation
	\begin{align}
	\begin{split}
	i\hbar \frac{\partial \mathcal{F}}{\partial t}
	=    
	\sum_{j=0}^{\mathcal{N}-1}\Bigg \langle \left(\prod_{i=1}^{\mathcal{N}-j}\hat\psi^{'\dagger}_{\mathcal{N}+1-i}\right)  \left( \hat A_{j} \overrightarrow{h}_{j+1}-\overleftarrow{h}_{j+1}^{'} \hat A_{j}\right) 
	\left(\prod_{i=j+1}^{\mathcal{N}}\hat\psi_{i}\right)   \Bigg \rangle  
	- i \hbar \, \mathcal{F} \sum_{i=1}^{\mathcal{N}}\frac{\gamma(\mathbf{r}_i,t)+\gamma(\mathbf{r}'_i,t)}{2}.
	\end{split}
	\end{align}
	%
%
The $\mathcal{N}$-point function function $\mathcal{F}$ is here evaluated using Schr\"odinger's non-evolving field operators. We can rewrite the above equation as 
%
%
\begin{align}\label{NpointEvol1}
\begin{split}
i\hbar \frac{\partial \mathcal{F}}{\partial t}
= &   
\sum_{j=0}^{\mathcal{N}-1}\Bigg \langle \left(\prod_{i=1}^{\mathcal{N}}\hat\psi^{'\dagger}_{\mathcal{N}+1-i}\right)  \left(\prod_{i=1}^{j}\hat\psi_{i}\right)  \overrightarrow{h}_{j+1}  \left(\prod_{i=j+1}^{\mathcal{N}}\hat\psi_{i}\right)   \Bigg \rangle \\
 & \qquad - \sum_{j=0}^{\mathcal{N}-1}\Bigg \langle \left(\prod_{i=1}^{\mathcal{N}-j}\hat\psi^{'\dagger}_{\mathcal{N}+1-i}\right)   \overleftarrow{h}_{j+1}^{'}  \left(\prod_{i=1}^{j}\hat\psi^{'\dagger}_{i}\right)  \left(\prod_{i=1}^{\mathcal{N}}\hat\psi_{i}\right)   \Bigg \rangle   - i \hbar \, \mathcal{F} \sum_{i=1}^{\mathcal{N}}\frac{\gamma(\mathbf{r}_i,t)+\gamma(\mathbf{r}'_i,t)}{2} ,
\end{split}
\end{align}
\end{strip}
where the field operators are explicitly given by $\hat\psi_{j} \equiv \hat\psi(\mathbf{r}_j)$ and $\hat\psi_{j}^{'\dagger} \equiv \hat\psi^\dagger(\mathbf{r}_j')$.
The evolution of $\mathcal{N}$-point functions can reproduce the evolution of basic quantities such as the density, the atom number, and higher-order correlations, thereby giving some concrete interpretation for the $\mathcal{N}$-point function. Eq.~\eqref{NpointEvol1} represents an equation of motion for various correlations of fermionic systems. The same equation describes the behavior of bosonic systems. Such an equation therefore unifies the dynamics of bosons and fermions as far as correlation functions are concerned.
%

\subsection{Mapping identity for two different evolutions of $\mathcal{N}$-point functions}

While the $\mathcal{N}$-point function above evolves in the Schr\"odinger picture, the useful exact mapping for closed many-body systems is expressed in terms of time-evolving quantum-field operators, \textit{i.e.}, in the Heisenberg picture. In order to take advantage of the mapping, we thus have to write the equation of motion of the Schr\"odinger $\mathcal{N}$-point function into the Heisenberg-picture form. One would think that it suffices to replace the density operator by an evolving field operator in order to write the Lindblad equation into the Heisenberg picture. Such a solution is quite naive for at least two reasons. The density matrix is not a standard quantum-mechanical operator, but a projective operator built out of state kets. Moreover the resulting equation fails to evolve the product of two operators properly, which is one fundamental way of testing the equations of motions of quantum mechanical operators. Thus the mapped system would be physically not realisable and consequently irrelevant.
In order to resolve the discrepancy, a Langevin force may be introduced into the problem, leading to the Heisenberg-Langevin equation. The price to pay when using such a counterpart of Lindblad equation is to deal with the stochasticity of the system, see the books~\cite{Zoller2004Book,Haken1970Book}. A cheaper and more experiment-oriented solution, however, consists in considering only the evolution of expectation values. The $\mathcal{N}$-point function encloses expectation values of various measurable physical quantities.


Considering that expectation values of given operators are unchanged irrespective of the quantum mechanical picture and using again the unitary transformation~\eqref{unitary}, the equation of motion of $\mathcal{N}$-point functions~\eqref{NpointEvol1} can be rewritten using the time-evolving quantum fields as
\begin{strip}
\begin{align}\label{NpointEvol2}
\begin{split}
\frac{\partial  \mathcal{F} }{\partial t}
= &    
\frac{1}{i\hbar} \sum_{j=0}^{\mathcal{N}-1}\Bigg \langle \left(\prod_{i=1}^{\mathcal{N}}\hat\psi^{'\dagger}_{\mathcal{N}+1-i}(t)\right)  \left(\prod_{i=1}^{j}\hat\psi_{i}(t)\right)  \overrightarrow{h}_{j+1}(t)  \prod_{i=j+1}^{\mathcal{N}}\hat\psi_{i}(t)   \Bigg \rangle 
\\
& \qquad
-\frac{1}{i\hbar} \sum_{j=0}^{\mathcal{N}-1}\Bigg \langle \left(\prod_{i=1}^{\mathcal{N}-j}\hat\psi^{'\dagger}_{\mathcal{N}+1-i}(t)\right)   \overleftarrow{h}_{j+1}^{'}(t)  \left(\prod_{i=1}^{j}\hat\psi^{'\dagger}_{i}(t)\right)  \prod_{i=1}^{\mathcal{N}}\hat\psi_{i}(t)    \Bigg \rangle  
- \mathcal{F} \sum_{i=1}^{\mathcal{N}}\frac{\gamma(\mathbf{r}_i,t)+\gamma( \mathbf{r}_i',t)}{2} ,
\end{split}
\end{align}
\end{strip}
where $\hat\psi_{j}(t) \equiv \hat\psi(\mathbf{r}_j,t)$,
$\hat{\psi}_{j}^{'\dagger}(t) \equiv \hat{\psi}^\dagger(\mathbf{r}_j',t)$, 
$\overrightarrow{h}_{j}(t) \equiv h(\mathbf{r}_{j},t)$, and 
$\overleftarrow{h}_{j}'(t) \equiv h(\mathbf{r}_{j}',t)$, with
\begin{align}
\begin{split}\label{ham1}
h(\mathbf{r},t)=  & - \frac{\hbar^2}{2M}  \nabla_{\mathbf{r}}^2  + V (\mathbf{r},t)   	\\
& +
\int d^{D}\mathbf{r}''  \,  \hat{\psi}^{\dagger} (\mathbf{r}'',t)  U(\mathbf{r}'',\mathbf{r},t)     {\hat{\psi}} (\mathbf{r}'',t) .
\end{split}
\end{align}
The difference between Eqs.~\eqref{NpointEvol1} and \eqref{NpointEvol2} is that the operators in the latter are expressed in the Heisenberg picture, so that we can have $h(\mathbf{r},t) \hat{\psi}(\mathbf{r},t) \equiv i\hbar \, \partial \hat{\psi}(\mathbf{r},t) /\partial t$ in the absence of dissipation, which is simply the Heisenberg equation. Likewise, the operator $h(\mathbf{r},t)$ in Eq.~\eqref{ham1} is the Heisenberg picture equivalent of $h(\mathbf{r})$ in Eq.~\eqref{ham0}. Like their Schr\"odinger picture counterparts, the operators $\overrightarrow{h}_{j}(t)$ and $\overleftarrow{h}_{j}^{'}(t)$ still apply only on $\hat\psi_{j}(t)$ and $\hat\psi^{'\dagger}_{j}(t)$, respectively, in the direction indicated by the arrow.


Eq.~\eqref{NpointEvol2} may describe the time evolution of expectation values of various physical quantities in a given experiment, say A. Using the mapping~\eqref{H.mapping}, we can show that the corresponding expectation values in another experiment, B, still obey a similar equation with potentials and dissipation rate different but appropriately related to those in A. 
A straightforward calculation leads to the Lindblad evolution of the mapped $\mathcal{N}$-point function~\eqref{H.mapping2}. Such an evolution is governed by an equation similar to Eq.~\eqref{NpointEvol2}, with the substitutions 
\begin{align}
\begin{split}
\hat\psi_{j}  & \to  \hat\Psi(\lambda \mathbf{r}_j,\tau), ~ \hat{\psi}_{j}^{'\dagger}  \to \hat{\Psi}^\dagger(\lambda \mathbf{r}_j',\tau), \\
h_{j} & \to \tilde h(\lambda \mathbf{r}_{j},\tau), ~ \tilde h(\mathbf{r},t)  =
 - \frac{\hbar^2}{2M}  \nabla_{\mathbf{r}}^2 + \tilde V (\mathbf{r},t)   	\\
& \qquad \qquad +
\int d^{D}\mathbf{r}''  \,  \hat{\Psi}^{\dagger} (\mathbf{r}'',t) \tilde U(\mathbf{r}'',\mathbf{r},t)     {\hat{\Psi}} (\mathbf{r}'',t) .
\end{split}
\end{align}
More importantly, the corresponding mapping identity for $\mathcal{N}$-point functions is the same as in Eq.~\eqref{H.mapping2}, but in addition a mapping identity for local dissipation rates has to be taken into account, which reads
\begin{align}\label{H.mapping3}
\begin{split}
\gamma(\mathbf{r}_i,t) & \to \tilde\gamma( \mathbf{r}_i, t)= \lambda(t)^2 \gamma(\lambda\mathbf{r}_i,\tau(t)).
\end{split}
\end{align}

Thus the $\mathcal{N}$-point functions $\mathcal{F}$ and $\tilde{\mathcal{F}}$ which are related in accordance with the quantum-field mapping, both satisfy the same equation~\eqref{NpointEvol2} derived from the Lindblad master equation~\eqref{Lindblad}. Therefore, for a given experiment A with a dissipative quantum gas described by the set $\{\psi, U, V, \gamma, \mathcal{F} \}$, an equivalent but different experiment B described by the set $\{\Psi, \tilde{U}, \tilde{V}, \tilde{\gamma}, \tilde{\mathcal{F}} \}$ will exist, such that both experiments can be exactly mapped onto each other. The mapping identities are given by Eqs.~\eqref{H.mapping}, \eqref{H.mapping1}, \eqref{H.mapping2}, and \eqref{H.mapping3}. This makes our exact space-time mapping relevant  for both non-dissipative and dissipative quantum systems. The $\mathcal{N}$-point functions can capture any observables, including the $\mathcal{N}$-body non-local correlation functions~\cite{Nandani2016,Piroli2016}, and the hydrodynamic fields~\cite{Henkel2017} which are the mean density $\langle\hat{n}\rangle=\langle \hat{\psi}^\dagger(\mathbf{r})\hat{\psi}(\mathbf{r}) \rangle$, the field correlations $G_1(\mathbf{r}-\mathbf{r}')=\langle \hat{\psi}^\dagger(\mathbf{r}')\hat{\psi}(\mathbf{r}) \rangle$, the density correlations $C(\mathbf{r}-\mathbf{r}')=\langle \hat{n}(\mathbf{r}')\hat{n}(\mathbf{r}) \rangle - \langle\hat{n}\rangle^2$, and the density-density correlation $G_2(\mathbf{r}-\mathbf{r}')=\langle (\hat{n}(\mathbf{r}')-\langle\hat{n}\rangle)(\hat{n}(\mathbf{r})-\langle\hat{n}\rangle) \rangle/\langle\hat{n}\rangle$~\cite{Hung2011,Naraschewski1999}. 

\section{A concrete example of mapping between two experimental situations} \label{secB}

In this section, we describe two possibly achievable experiments with bosons, say A and B. By using numerical experiments we show how a spacetime mapping can be used to mimick the evolution of correlations in B from the direct evolution of correlations measured in A. More clearly, the wording 'experiments' we use does not actually mean achieved experiments, since we have performed no experimental works. It simply denotes a realistic situation that can be experimentally explored or reproduced in a quantum gas laboratory. Data for the mapping are here given by the condensate wave function, which is the simplest example we could take for illustrating our results. Such a wave function is generated through numerical computations of the imaginary Gross-Pitaevskii (iGP) equation. While the iGP equation does not directly stem from Eq.~\eqref{NpointEvol1}, we show in the Appendix how it can be derived directly in a similar way. There we also show how its derivation can follow from Eq.~\eqref{NpointEvol1}.

\subsection{Description of the experimental situations}

We consider two potential experiments with simple lossy quantum systems made of a harmonically trapped and elongated ultracold gas of bosons with repulsive interaction onto which is acting a highly controlled dissipation. In experiment A, the gas with constant interparticle interaction is confined in a harmonic trapping potential $V_A=M\omega_A^2 r^2/2$ and the amplitude of local dissipation rate is modulated in time. We achieve a corresponding experiment B with trapping potential $V_B=M\omega_B^2 r^2/2$ and modulated interparticle interaction strength by choosing
\begin{equation}
\lambda(t) = \frac{1}{\sqrt{a \cos(2\omega t) + b}}, 
\end{equation}
where $a = (1-\eta^2)/2$ and $b = (1+\eta^2)/2$, and we take $\eta$ to be the ratio of trapping frequencies in both experiments, $\eta = \omega_A/\omega_B$, and $\omega \equiv \omega_B$. The free parameter satisfies $\lambda(0)=1$ and $\dot{\lambda}(0)=0$, and shows how to modulate the dissipation and interaction strengths in experiments A and B. The time $t_A$ when the data are measured in experiment A can be mapped onto the time $t_B$ when the corresponding data are obtained in experiment B, such that 
\begin{equation}\label{TimeMap}
t_A = \int_0^{t_B} \lambda(t)^2 dt.
\end{equation}
If we take for instance $t_B=20\pi$, we obtain $t_A \approx 41.89$.
Even though the present setting is not very general, as it does not allow to independently tune the strength and modulation frequency of the interparticle interaction, the scenario we describe here is experimentally achievable. For a contact interaction ($s = D$), the modulation of interparticle interactions can be realized with a time-dependent magnetic field via Feshbach resonances~\cite{Chin2010,Makotyn2014}. 
Experimentally it has been possible to engineer tunable dissipations in quantum systems using a technique called scanning electron microscopy. The working principle of this electron microscopy for quantum gases consists in scanning a focused electron beam over the atom cloud. The electron beam is shot onto the condensate, and the collision of electrons with the trapped atom cloud locally causes the ionization of single atoms turning them onto untrapped ions that escape from the condensate~\cite{BarontiniPRL2013}. The fleeing ions can then be subsequently detected by an ion detector~\cite{GerickePhD}. The electron beam creates a fully controllable and environmentally induced imaginary potential acting on the ultracold quantum gas.

Theoretically, an ultracold quantum gas with a source or sink of particles is best described by the Lindblad master equation. In the mean-field regime, however, such a system may be well described by a Gross-Pitaevskii equation with an imaginary potential where contact interaction is assumed~\cite{BarontiniPRL2013}. One of the reasons the Gross-Pitaevskii equation is preferred to the Lindblad equation is easier numerical treatment since the Gross-Pitaevskii equation is in general much more tractable in the limit of large atom numbers where solving more accurate equations is hard and time-consuming. Without loss of generality, we consider the system to be one-dimensional. In the dimensionless form, the Gross-Pitaevskii equation with imaginary potential can be written as follows~\cite{Gericke2008,Wamba2011,Zezyulin2016}:
\begin{equation}\label{GPE}
i \frac{\partial \phi}{\partial t} = -\frac{1}{2} \frac{\partial^2 \phi}{\partial x^2} + V(x)\phi + g(t) \vert \phi\vert^2 \phi + i\gamma(x,t) \psi .
\end{equation}
The function $\phi$ depends on time $t$ and space $x$, and represents the condensate wave function. It corresponds to the field operator $\hat{\psi}$, i.e., $\langle \hat{\psi} \rangle \equiv \phi$. The coefficient $g(t)$ is the strength of interaction between the gas particles. The condensate is harmonically trapped such that $V(x)=\omega^2 x^2/2$.  The prefactor $\gamma(x,t)$ is the local dissipation rate given by~\cite{BarontiniPRL2013,Gericke2008}:
\begin{equation}
\gamma(x,t) = \sigma(t) e^{-\frac{x^2}{2 w(t)^2}},
\end{equation}
where $\sigma \propto I/w^2$, with $I(t)$ being the electron current and $w(t)$ the waist of the beam which are both tunable independently. A similar dissipation rate produced by a cylindrically focused laser beam was used to create a single Gaussian defect in the study of dissipative transport of Bose-Einstein condensates~\cite{Hulet2010}.

In what follows, we will numerically solve the Gross-Pitaevskii equation above for a Bose-Einstein condensate with imaginary potential to examine the dynamics of the system for the two experiments A and B. In the numerical experiment A, the interaction strength $g_A(t)=g_0$ and the beam waist $w_A(t)=w_0$ are constant. The amplitude of the loss rate is modulated in time in such a way that $\sigma_A(t)=\sigma_0/ \lambda(t)^2$. In the numerical experiment B, the waist of the beam as well as the interaction strength are modulated in time such that $w_B(t) = w_0 / \lambda(t)$, and $g_B(t) = g_0 \lambda(t)$. The amplitude of the loss rate $\sigma_B(t)=\sigma_0$ is constant. Thus the amplitude and waist of the beam is differently modulated in both experiments. With the above choice of parameters, experiments A and B may be described by the quantum fields $\hat{\psi}$ and $\hat{\Psi}$, respectively, and then can be related onto each other through our mapping. 
We perform the numerical analysis based on the split-operator method including the fast Fourier transform. The initial state will be prepared by relaxation from a condensate background within the Thomas-Fermi regime.

\subsection{Mean-field mapping of the experiments}

We consider the case where the beam is narrow, such that the average waist is less than a quarter of the initial condensate radius. Only the particles located around the trap center are then kicked out of the condensate. We display the results from direct computations of the Gross-Pitaevskii equation for both experiments, and then concretely show how the result of one experiment can be recovered from the other by using the mapping.

\subsubsection{Direct evolutions of the system in both experiments}
In order to depict the evolutions of the system, we may compute the time evolution of any dynamical variables like the condensate radius, the density, or the volume integrated local $\nu$-body correlation function defined as~\cite{Naraschewski1999,Nandani2016,Armijo2010,Gangardt2003}
\begin{equation}
g^{(\nu)}(t) \propto \int \vert \phi(x,t) \vert^{2(\nu+1)} dx,
\end{equation}
where $\nu=0,1,2,3$ correspond to the density, and local one-, two- and three-body correlation functions, respectively.

In Fig.~\ref{Fig1}, we portray the densities $|\phi_A(x,t)|^2$ and $|\phi_B(x,t)|^2$ for both experiments A and B, respectively. In experiment A, the only time-dependent parameter is the amplitude of the loss rate, and thus only excitations with small amplitude form on top of the cloud as shown in Fig.~\ref{Fig1}(a). In experiment B both the beam waist and the interparticle interaction strength are nontrivially changing with time in a periodic way. The condensate strongly breathes as its radius and maximum density vary periodically in time, see Fig.~\ref{Fig1}(b). It is clear that experiments A and B are very different, but as we will show in what follows, they can be exactly mapped onto each other using our mapping. 

\subsubsection{Using the mapping to mimick the evolution of B from the direct evolution of A}
\begin{figure}[tbh]
\centering
\includegraphics[scale=0.197]{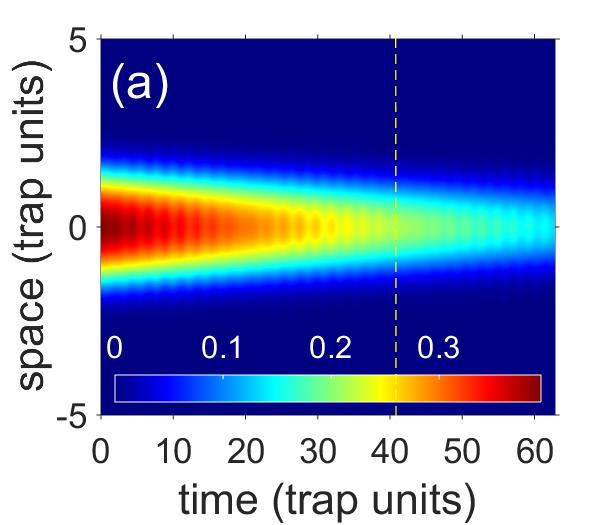}
\includegraphics[scale=0.197]{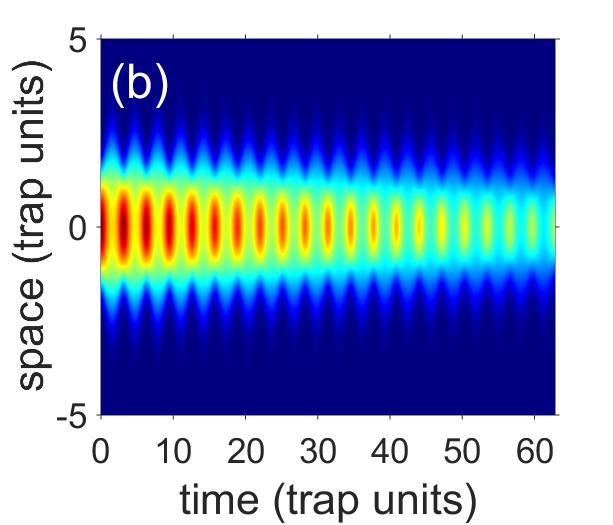}
\caption{\footnotesize (Color online) Density evolution computed directly using the Gross-Pitaevskii equation~\eqref{GPE} for times up to $t=20\pi$ in experimental situations A (a) and B (b). The dashed line in panel (a) depicts the time $t_A \approx 41.9 \equiv t^*$ which may be mapped onto $t_B=20\pi$ according to Eq.~\eqref{TimeMap}. The values of simulation parameters used are $\omega_A=1.5$, $\omega_B=1.0$, $g_0=10.0$, $\sigma_0=3.985\times 10^{-2}$, and $w_0=0.1563$.} \label{Fig1}
\end{figure}

\begin{figure}[tbh]
\includegraphics[scale=0.197]{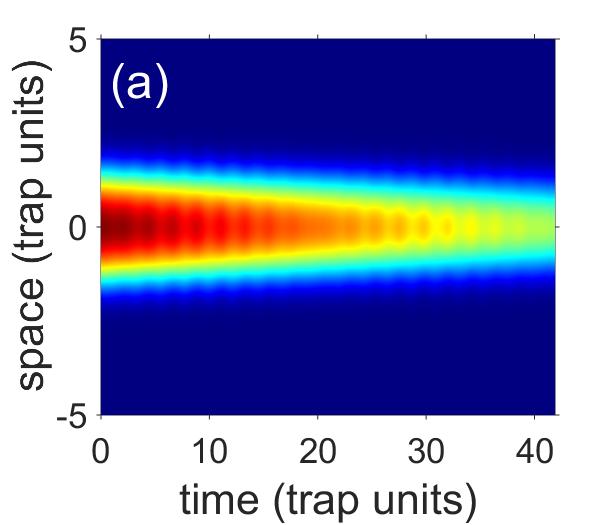}
\includegraphics[scale=0.197]{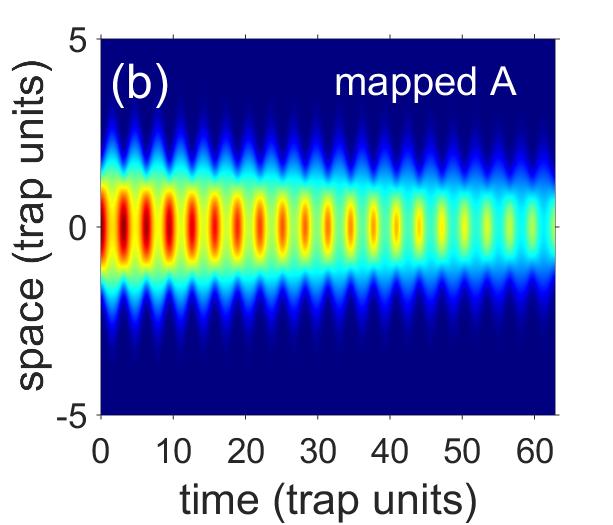}
\includegraphics[scale=0.197]{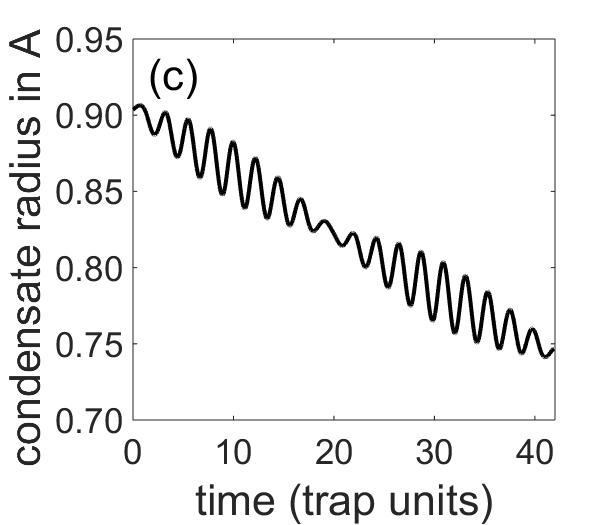}
\includegraphics[scale=0.197]{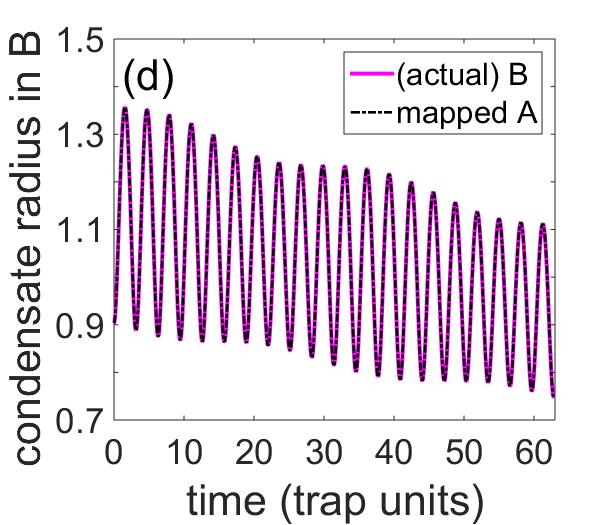}
\includegraphics[scale=0.197]{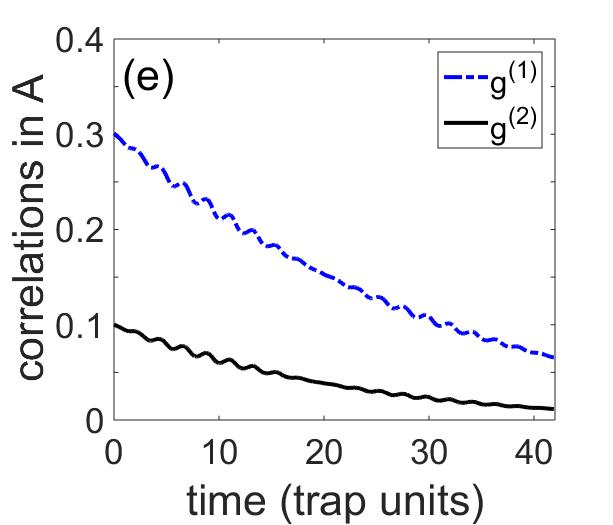}
\includegraphics[scale=0.197]{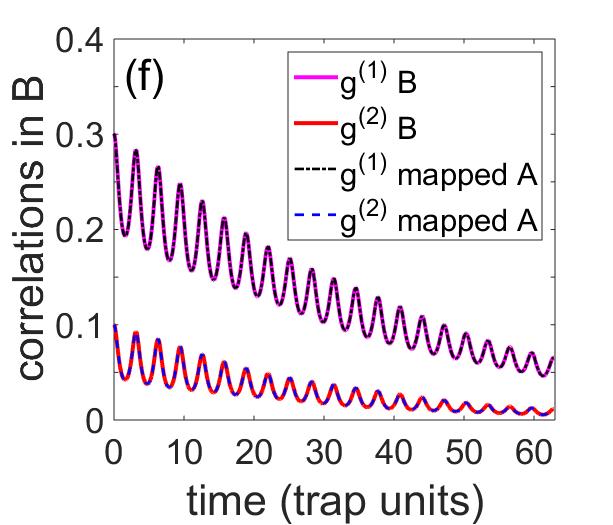}
\caption{\footnotesize (Color online) Exactly mapping the data in the experimental situation B from the data in the experimental situation A. The left panels (a), (c) and (e) show the evolution of the density, condensate radius and local correlations in experiment A for runs up to time $t^*$, respectively. The right panels (b), (d), (f) depict a comparison between the data from the direct numerical experiment B and the result mapped~A obtained by mapping the data from the direct numerical experiment A, through the density, condensate radius and local correlations, respectively.}\label{Fig2}
\end{figure}

Suppose the data from experiment A as portrayed in Fig.~\ref{Fig1}(a) are described by the field operator $\hat{\psi}$ or the condensate wave function $\phi_A$. We reconsider those data for times $t \in [0, t^*]$ where $t^* \approx 41.9$, and we display the density, the condensate radius, and the correlators $g^{(1)}$ and $g^{(2)}$ in Figs.~\ref{Fig2}(a), (c), (e), respectively. Using these data, we can deduce the corresponding data of experiment B as portrayed in Fig.~\ref{Fig1}(b) without running it. For this, we introduce the data in the mapping~\eqref{H.mapping1}, and obtain the data described by the field operator $\hat{\Psi}$ or condensate wave function $\phi_B$ (data labelled in the plots by $\mathrm{mapped~A}$), which we display in Figs.~\ref{Fig2}(b), (d), (f) comparatively with data from Fig.~\ref{Fig1}(b) for times $t \in [0, 20\pi]$. We readily realize that the density evolution in Fig.~\ref{Fig2}(b) is exactly identical to the one in Fig.~\ref{Fig1}(b). Moreover, there is a perfect overlap between the condensate radii and local correlations in Figs.~\ref{Fig2}(d), (f). Thus the data $\mathrm{mapped~A}$ obtained by applying the mapping to A are nothing but the data directly obtained from experiment B. Hence our mapping clearly allows to exactly mimick the evolution of B from the direct evolution of A.
Conversely, it is possible to get the data in Fig.~\ref{Fig1}(a) from those in Fig.~\ref{Fig1}(b). In that case, we have to inverse the relation~\eqref{H.mapping1} in such a way that the field operator $\hat{\psi}$ is expressed as a function of the field operator $\hat{\Psi}$. The same mapping holds for any possible correlators or experimental data that can be measured in both experiments.

\subsubsection{Special cases}
The mapping may take a particularly significant meaning in some particular regimes.
\paragraph*{\textit{Very large beam waist.}} We consider a regime when the electron beam is large enough to hit the condensate anywhere at a time. Because the waist is very large (we take e.g., $w_0=15.63$, keeping $\sigma_0=7.970 \times 10^{-3}$), the local dissipation rate reduces to $\gamma(x,t) \approx \sigma(t)$ which is space-independent. Then the mapping transforms a time periodic modulation of the dissipation rate $\gamma(t)$ onto a time periodic modulation of the interaction strength $g(t)$ between the particles of the dissipative quantum gas. Dissipation from an electron beam can then be used to control the interaction strength between the particles of a quantum gas.
It is true that controlling the interparticle interaction in many types of quantum gas species is usually done via magnetic, optical, and magneto-optical Feshbach resonances. The specific type of Feshbach resonance is chosen based on the type of particle, and is consequently dependent on the resonance width, which turns out to be very narrow for some species. Therefore, Feshbach resonance technique may not be efficient in some regimes, for instance, due to the narrowness of the resonance width or a faster depletion of the quantum gas sample through recombination processes.
In such regimes, our method can be appropriate.
\paragraph*{\textit{Weak interaction regime.}} In the limit when $g_0$ tends to zero, the quantum gas is close to an ideal gas and the effect of dissipation can be studied alone. The mapping then transforms a time periodic modulation of the dissipation rate amplitude $\sigma(t)$ onto a time periodic modulation of the beam waist $w(t)$. In this case, the mapping helps matching the effect on the condensate of changing the electron current for fixed beam waist and the effect of changing the beam waist at constant electron current. This regime can also well describe the behavior of polaritons and photon condensates where the interparticle interaction is very weak in general~\cite{Milan2018,Stein2019}. It should be noted that the mapping is still valid even when mean field breaks down, since it is orginally formulated in terms of quantum fields.

\section{Conclusion} \label{concl}

In this paper, we have presented an exact spacetime mapping of two experimental situations with open quantum systems in a regime of loss or gain. The mapping is formulated with the help of $\mathcal{N}$-point functions of two evolutions of open quantum gases that obey the Lindblad equation. Our formalism is general as it extends a previous result on quantum field mapping for closed systems~\cite{WPA1} to the general case of Markovian open quantum systems with gain and/or loss.
For this, we have considered an open many-body quantum system consisting of a quantum gas in any arbitrary dimension that interacts with the environment and evolves in the regime when the Lindblad generators feature a loss or a gain. Given that expectation values are unchanged in different pictures of quantum mechanics, we have used the evolution of quantum fields that describe the gas dynamics to derive the Heisenberg evolution of any arbitrary $\mathcal{N}$-point correlation function of the system.
Our previous result on the quantum field mapping for closed quantum systems is rewritten in terms of wave functions, and the extension to open quantum systems is achieved by relating onto each other the $\mathcal{N}$-point functions of two different evolutions of the open quantum system. Such a result is valid for both bosonic and fermionic systems.
Because the imaginary Gross-Pitaevskii equation is a mean-field approximation and a special case of the Lindblad master equation for lossy bosonic systems that can provide quantitative results, an illustrative example of the mapping has been provided based on such an equation. That example includes two achievable experiments which are the modulation of the local dissipation rate and the modulation of the interaction strength in ultracold gases of bosons confined in harmonic traps.
The result suggests that a modulation of the dissipation rate can be used to control the interaction between particles in open quantum many-body systems.
Our result is a general theoretical model that can be useful for mimicking quantum gas experiments with dissipation and for benchmarking relevant theoretical approximations, especially in challenging regimes of dissipation that make the system hardly tractable.
In our future works, it may be worth trying to further generalize our mapping to explicitly describe autocorrelation functions and possibly non-Markovian open quantum systems.

\section*{Acknowledgements}
Part of this work was done during a research visit of EW at the Technical University Kaiserslautern, Germany. EW thanks his host Prof.~Dr. James R. Anglin for that opportunity and for deep and fruitful discussions thereof. EW acknowledges financial support from the Alexander von Humboldt Foundation, and from the Simons Associateship joint programme of the Abdus Salam International Center for Theoretical Physics and of the Simons Foundation. The work is also funded by the Deutsche Forschungsgemeinschaft (DFG, German Research Foundation) under the project number 277625399 - TRR 185.


\bigskip

\setcounter{equation}{0}
\setcounter{figure}{0}
 \renewcommand{\theequation}{A-\arabic{equation}}
 \renewcommand{\thefigure}{A-\arabic{figure}}

 \section*{Appendix: The imaginary Gross-Pitaevskii equation}

As shown in the main text, this paper essentially focuses on the evolution of $\mathcal{N}$-point functions, which have an equal number of creation and annihilation operators.  In section~\ref{secB}, however, we discussed as a bosonic example the condensate wave function, which is the expectation value of only one creation or annihilation operator governed by the imaginary Gross-Pitaevskii (iGP) equation.
It should be noted that the analogue of the iGP equation does not exist for fermionic systems since the expectation value of any field operator would be zero. This is a consequence of the macroscopic occupation of the lowest energy state being guaranteed for bosons but forbiden for fermions due to the Pauli exclusion principle.
In this Appendix, we discuss the derivation of the iGP equation and the equation that rules the condensate decay rate, i.e., the particle loss rate.
First, the derivation of the iGP equation is carried out directly following the method presented in section~\ref{LindSec} for a simple example where $\hat{\mathcal{O}}=\hat{\psi}$. Secondly, the iGP equation is retrieved from the evolution of the $1$-point function, and then the equation that governs the condensate decay rate is deduced from such an evolution.

\subsection*{A.1- Direct derivation}

Suppose we have $\hat{\mathcal{O}}=\hat{\psi}$.  
Let us expand separately the two terms in the right-hand side of Eq.~\eqref{EqA2}. Then the first one reads
\begin{strip}
\begin{align} 
\begin{split}
\left \langle \left[\hat\Psi ,\hat H   \right]_{-} \right \rangle   =&
\Big \langle   \int d^{D}\mathbf{r}'\,   	\left[\hat{\Psi} , \hat{\Psi}^{'\dagger}    
\left(  - \frac{\hbar^2}{2M}  \nabla_{\mathbf{r}'}^2 +V (\mathbf{r}',t)    \right)  \hat{\Psi}' \right]_{-} 
	+
\frac{1}{2}   \int d^{D}\mathbf{r}'  d^{D}\mathbf{r}'' \,  \left[\hat{\Psi}, \hat{\Psi}^{' \dagger} \hat{\Psi}^{''\dagger}   U(\mathbf{r}',\mathbf{r}'',t)   {\hat{\Psi}} ^{''} {\hat{\Psi}} ^{'} \right]_{-} \Big \rangle \\
=&
\Big \langle    
\left(  - \frac{\hbar^2}{2M}  \nabla_{\mathbf{r}}^2 +V (\mathbf{r},t)    \right)  \hat{\Psi}(\mathbf{r}) 	+
\int d^{D}\mathbf{r}'  \,  \hat{\Psi}^{\dagger} (\mathbf{r}')   U(\mathbf{r}',\mathbf{r},t)    {\hat{\Psi}} (\mathbf{r}') {\hat{\Psi}}(\mathbf{r})  \Big \rangle = \langle h(\mathbf{r}) {\hat{\Psi}}(\mathbf{r}) \rangle . 
\end{split}
\end{align}
\end{strip}
We have used commutator identities, the boson canonical commutators as well as the relation $[\hat{A},\hat{B}]_{-}=[\hat{A},\hat{A}^\dagger]_{-} \partial \hat{B}/ \partial{\hat{A}^\dagger} $, or equally the $-$ branch of the rule~\eqref{ABC}. 
The second term of Eq.~\eqref{EqA2} yields
\begin{align} 
\begin{split}
 {\textrm Tr}\left(\hat\Psi  \mathcal{\hat L}\hat\rho \right)   
& = -  \int d^{D}\mathbf{r}'\, \frac{\gamma(\mathbf{r}',t)}{2}   {\textrm Tr}\Big(\hat\Psi  \Big( \hat\Psi^{'\dagger}\hat\Psi' \hat\rho  
\\
& \qquad \qquad +\hat\rho \hat\Psi^{'\dagger}\hat\Psi'- 2\hat{\Psi}' \hat{\rho}\hat{\Psi}^{'\dagger} \Big) \Big) 
\\
& = -  \int d^{D}\mathbf{r}'\, \frac{\gamma(\mathbf{r}',t)}{2}   {\textrm Tr}\left( [\hat\Psi , \hat\Psi^{'\dagger}]\hat\Psi' \hat\rho     \right) \\
& +  \int d^{D}\mathbf{r}'\, \frac{\gamma(\mathbf{r}',t)}{2}   {\textrm Tr}\left( \hat\Psi^{'\dagger}[\hat\Psi , \hat\Psi'] \hat\rho     \right) .
\end{split}
\end{align}
Using the boson canonical commutator as well as the rule $\langle\hat{\mathcal{O}}\rangle = \mathrm{Tr}\left( \hat{\mathcal{O}} \hat{\rho}\right)$, we get
\begin{align} 
\begin{split}
 {\textrm Tr}\left(\hat\Psi  \mathcal{\hat L}\hat\rho \right)  
=&
-   \int d^{D}\mathbf{r}'\, \frac{\gamma(\mathbf{r}',t)}{2}   \Big \langle  [\hat\Psi , \hat\Psi^{'\dagger}]\hat\Psi' \Big \rangle   
\\
=&
-  \int d^{D}\mathbf{r}'\, \delta^D(\mathbf{r} - \mathbf{r}'  )    \frac{\gamma(\mathbf{r}',t)}{2}  \langle \hat{\Psi}'   \rangle    
\\
=&
-  \frac{\gamma(\mathbf{r},t)}{2}    \langle \hat{\Psi}  \rangle. 
\end{split}
\end{align}
Summing together both terms, we therefore obtain
\begin{align}
\begin{split}
i\hbar 		\frac{\partial}{\partial t}\langle \hat\Psi \rangle 
= &  \Big \langle    
\left(  - \frac{\hbar^2}{2M}  \nabla_{\mathbf{r}}^2 +V (\mathbf{r},t)    \right)  \hat{\Psi}(\mathbf{r}) 
\\	
& \qquad +
\int d^{D}\mathbf{r}'  \,  \hat{\Psi}^{'\dagger} U(\mathbf{r}',\mathbf{r},t)   \hat{\Psi} \hat{\Psi}' \Big \rangle
\\
& \qquad - i  \hbar  \frac{\gamma(\mathbf{r},t)}{2} \langle \hat{\Psi} \rangle 
\\
& \equiv \langle \overrightarrow{h}(\mathbf{r}) \hat{\Psi}(\mathbf{r}) \rangle - i  \hbar  \frac{\gamma(\mathbf{r},t)}{2}   \langle \hat\Psi(\mathbf{r})  \rangle .
\end{split}
\end{align}
In this equation, the correlation function $\langle \hat\Psi \rangle$ represents the condensate wave function.

\paragraph*{{\textit Mean-field illustration.}} In the mean-field regime, expectation values can readily be factorized, and we have
\begin{align}\label{A6}
\begin{split}
i\hbar 	\frac{\partial}{\partial t}\langle \hat{\Psi} \rangle 
=&     
\left(  - \frac{\hbar^2}{2M}  \nabla_{\mathbf{r}}^2 +V (\mathbf{r},t)    \right) \langle \hat{\Psi} \rangle 
\\
&+\int d^{D}\mathbf{r}'  \,  \langle\hat{\Psi}^{'\dagger} \rangle  U(\mathbf{r}',\mathbf{r},t) \langle  \hat{\Psi}\rangle \langle \hat{\Psi}' \rangle 
\\
& - i  \hbar  \frac{\gamma(\mathbf{r},t)}{2}   \langle \hat{\Psi} \rangle.
\end{split}
\end{align}
We set for the mean-field wave functions to be
\begin{align}
\begin{split}
\hat\Psi (\mathbf{r}) \approx \langle \hat\Psi (\mathbf{r})  \rangle (t)  =  \phi(\mathbf{r},t ), \\
\hat\Psi ^{\dagger}(\mathbf{r}) \approx \langle \hat\Psi ^{\dagger}(\mathbf{r})  \rangle (t)  =  \phi^*(\mathbf{r},t ) ,
\end{split}
\end{align}
which are expectation values of time independent operators evaluated at time $t$, and \textit{not} the expectation values of time-evolving operators. The quantity $\phi(\mathbf{r},t )$ is actually the (quasi-)condensate wave function. 
Thus from \eqref{A6} we finally get
\begin{align}
\begin{split}
i\hbar \frac{\partial \phi(\mathbf{r},t ) }{\partial t} 
=  
\Big(  - \frac{\hbar^2}{2M}  \nabla_{\mathbf{r}}^2 &+V (\mathbf{r},t)  	
\\
+\int d^{D}\mathbf{r}'  \,  \phi^{\ast} (\mathbf{r}',t )  & U(\mathbf{r}',\mathbf{r},t) \phi(\mathbf{r}',t ) 
\\
&- i \hbar   \frac{\gamma(\mathbf{r},t)}{2}     \Big)    \phi(\mathbf{r},t )  .
\end{split}
\end{align}
In the case of hard-sphere contact potential, we have $U(\mathbf{r}',\mathbf{r},t)=g(t)\delta^D(\mathbf{r}-\mathbf{r}')$.  Thus 
\begin{align}\label{GPE2}
\begin{split}
i\hbar \frac{\partial \phi(\mathbf{r},t ) }{\partial t} 
&=  
\Big(  - \frac{\hbar^2}{2M}  \nabla_{\mathbf{r}}^2 +V (\mathbf{r},t)  
\\
&+g(t) \, \vert  \phi(\mathbf{r},t ) \vert^2
- i  \hbar  \frac{\gamma(\mathbf{r},t)}{2}     \Big)    \phi(\mathbf{r},t )  ,
\end{split}
\end{align}
which is a more general form of the Gross-Pitaevskii equation with imaginary potential as given in Eq.~\eqref{GPE} of the main text. A similar derivation was provided in Ref.~\cite{BarontiniPRL2013}, see its supplemental material.
By combining Eq.~\eqref{GPE2} and its conjugate, it is possible to derive the condensate decay rate. However we show here that the decay rate can be obtained directly using our evolution of a $1$-point function, which means that the iGP can be inversely derived from the $1$-point function.

\subsection*{A.2- Derivation from a $1$-point function}

Suppose we now have $\hat{\mathcal{O}} = \hat\Psi^\dagger\left(\mathbf{r}'_1\right)  \hat\Psi\left(\mathbf{r}_1\right)$, \textit{i.e.}, similar to a local density operator, and $\mathcal{F}$ is similar to a one-body local density. The time evolution of a one-point function can then be given by Eq.~\eqref{NpointEvol1} for $\mathcal{N}=1$, so we have
\begin{align}\label{OnePt}
\begin{split}
i\hbar \frac{\partial}{\partial t}\langle \hat{\Psi}^{'\dagger}_1 \hat{\Psi}_1 \rangle 
=\Big \langle    
\hat{\Psi}^{'\dagger}_1 \left( -\overleftarrow{h}(\mathbf{r}'_1)+\overrightarrow{h}(\mathbf{r}_1) \right) \hat{\Psi}_1\Big \rangle 
\\
- i\hbar    \frac{\gamma(\mathbf{r}'_1,t) + \gamma(\mathbf{r}_1,t)}{2}   \langle \hat{\Psi}^{'\dagger}_1 \hat{\Psi}_1 \rangle.
\end{split}
\end{align}
Using this evolution equation for expectation values, we can obtain the imaginary Gross-Pitaevskii equation and even the decay rate equation.

\subsubsection*{A.2.1- Evolution in the mean-field regime: the iGP equation}

Consider the evolution equation of the $1$-point function in Eq.~\eqref{OnePt} in a regime where the mean-field approximation is valid. Then we can factorize the expectation values of all operators. Because the multiplication of expectation values is commutative and considering the evolution of $\langle \hat{\Psi}_1 \rangle\equiv\phi(\mathbf{r}_1,t)$ and $\langle \hat{\Psi}^{'\dagger}_1 \rangle \equiv \phi^*(\mathbf{r}_1,t)$, we get the following set of two equations:
\begin{align}\label{OnePt2}
\begin{split}
i\hbar \frac{\partial \langle \hat{\Psi}_1 \rangle}{\partial t} 
=&   \overline{h}(\mathbf{r}_1) \langle \hat{\Psi}_1 \rangle  
- i\hbar    \frac{\gamma(\mathbf{r}_1,t)}{2} \langle \hat{\Psi}_1 \rangle \\
i\hbar \frac{\partial \langle \hat{\Psi}^{'\dagger}_1 \rangle}{\partial t} 
=&   - \overline{h}(\mathbf{r}'_1)\langle    
\hat{\Psi}^{'\dagger}_1\rangle - i\hbar    \frac{ \gamma(\mathbf{r}_1',t)}{2}  \langle \hat{\Psi}^{'\dagger}_1 \rangle  .
\end{split}
\end{align}
In that equation, we have
\begin{align}\label{hami}
\begin{split}
\overline{h}(\mathbf{r})=  -& \frac{\hbar^2}{2M}  \nabla_{\mathbf{r}}^2  +  V (\mathbf{r},t)   	\\
& +
\int d^{D}\mathbf{r}''  \, \langle \hat{\psi}^{\dagger} (\mathbf{r}'') \rangle U(\mathbf{r}'',\mathbf{r},t)     \langle {\hat{\psi}} (\mathbf{r}'') \rangle .
\end{split}
\end{align}
By taking the complex conjugate of the second equation in the set~\eqref{OnePt2}, we realize that the two equations are identical. In the case of hard-sphere contact potential, we readily retrieve the iGP equation~\eqref{GPE2}. 

\subsubsection*{A.2.2- Evolution in the equal position limit: the decay rate}
In order to get the particle loss rate from Eq.~\eqref{OnePt}, consider the special case when the creation and annihilation operators act at the same position, i.e., $\mathbf{r}'_1=\mathbf{r}_1 \equiv \mathbf{r}$. Then we get the dissipative and von Neumann terms to be $- i\hbar  \gamma(\mathbf{r},t) \, n(\mathbf{r},t) $ and $\Big \langle    
\hat{\Psi}^{\dagger} \left( -\overleftarrow{h}(\mathbf{r})+\overrightarrow{h}(\mathbf{r}) \right) \hat{\Psi}\Big \rangle$, respectively, where $n(\mathbf{r},t)=\Big \langle \hat\Psi^\dagger(\mathbf{r})  \hat\Psi(\mathbf{r})  \Big \rangle_t$. In the Hamiltonian part, both the single- and two-particle interactions terms vanish, and thus the above equation becomes 
\begin{align}\label{A17}
\begin{split}
i\hbar \frac{\partial}{\partial t}n(\mathbf{r},t) 
=&\frac{\hbar^2}{2 M}\left(     
\Big \langle\hat{\Psi}  \overrightarrow{\nabla}_{\mathbf{r}}^2 \hat{\Psi}^{\dagger}\Big \rangle^{*} - \Big \langle\hat{\Psi}^{\dagger}\overrightarrow{\nabla}_{\mathbf{r}}^2  \hat{\Psi}\Big \rangle \right) \\
&- i\hbar  \gamma(\mathbf{r},t)\, n(\mathbf{r},t) .
\end{split}
\end{align}
It should be noted that $\overrightarrow{h}-\overleftarrow{h}$ is not a null operator because of the quantum pressure resulting from the kinetic term.

\paragraph*{{\textit Mean-field illustration.}} If we invoke mean-field approximation in the above equation, 
it yields the continuity equation
\begin{align}
\begin{split}
i\hbar \frac{\partial }{\partial t}|\phi\left(\mathbf{r},t\right)|^2
=\frac{\hbar^2}{2 M}\left(     
\phi  {\nabla}_{\mathbf{r}}^2 \phi^{*} -  \phi^{*} {\nabla}_{\mathbf{r}}^2  {\phi}  \right) - i\hbar  \gamma(\mathbf{r},t) \, \vert \phi(\mathbf{r},t) \vert^2.
\end{split}
\end{align}
Summing over all possible $\mathbf{r}$ finally gives the relation:
\begin{align}
\begin{split}
\frac{\partial N }{\partial t} &= - \int d^D\mathbf{r}  \,     \gamma(\mathbf{r},t)  \vert  \phi(\mathbf{r},t ) \vert^2 . 
\end{split}
\end{align}
Obviously, the decay of atom number is due to the loss. In the case when the loss rate is constant, we get a purely exponential decay given by $N(t) = N_0 \exp(-\gamma t)$.
It should be noted that the mean-field approximation is not necessary for the derivation of the above equation of atom number decay. Indeed, directly integrating Eq.~\eqref{A17} over space $\mathbf{r}$ leads to the same result.

\end{document}